\documentclass[graybox,natbib,nosecnum]{svmult}
\bibpunct{(}{)}{;}{a}{}{,} 

\pdfoutput=1   

\usepackage{mathptmx}       
\usepackage{helvet}         
\usepackage{courier}        
\usepackage{type1cm}        
\usepackage{subcaption}
\usepackage{amsmath}

\usepackage{makeidx}         
\usepackage{graphicx}        
\usepackage{multicol}        
\usepackage[bottom]{footmisc}
\usepackage[normalem]{ulem}	
\usepackage[pdftex,pdfpagemode={UseOutlines},bookmarks,bookmarksopen,colorlinks,linkcolor={blue},citecolor={blue},urlcolor={red}]{hyperref}  

\usepackage{soul}   

\makeindex             

\begin{document}

\title*{The Rossiter--McLaughlin effect in Exoplanet Research}
\author{Amaury H.M.J. Triaud}
\institute{Amaury H.M.J. Triaud \at University of Birmingham, School of Physics \& Astronomy, Edgbaston, B15 2TT, Birmingham, United Kingdom, \email{A.Triaud@bham.ac.uk}
\at University of Cambridge, Institute of Astronomy, Madingley Road, CB3 0HA, Cambridge, United Kingdom, 
}
%
%
\maketitle

\abstract{The {Rossiter--McLaughlin effect} occurs during a planet's {transit}. It provides the main means of measuring the sky-projected {spin--orbit angle} between a planet's orbital plane, and its host star's equatorial plane. Observing the {Rossiter--McLaughlin effect} is now a near routine procedure. It is  an important element in the orbital characterisation of transiting exoplanets. Measurements of the {spin--orbit angle} have revealed a surprising diversity, far from the placid, Kantian and Laplacian ideals, whereby planets form, and remain, on orbital planes coincident with their star's equator. This chapter will review a short history of the {Rossiter--McLaughlin effect}, how it is modelled, and will summarise the current state of the field before describing other uses for a spectroscopic {transit}, and alternative methods of measuring the {spin--orbit angle}.
}

\section{Introduction }

\begin{figure*}
\includegraphics[width=\textwidth]{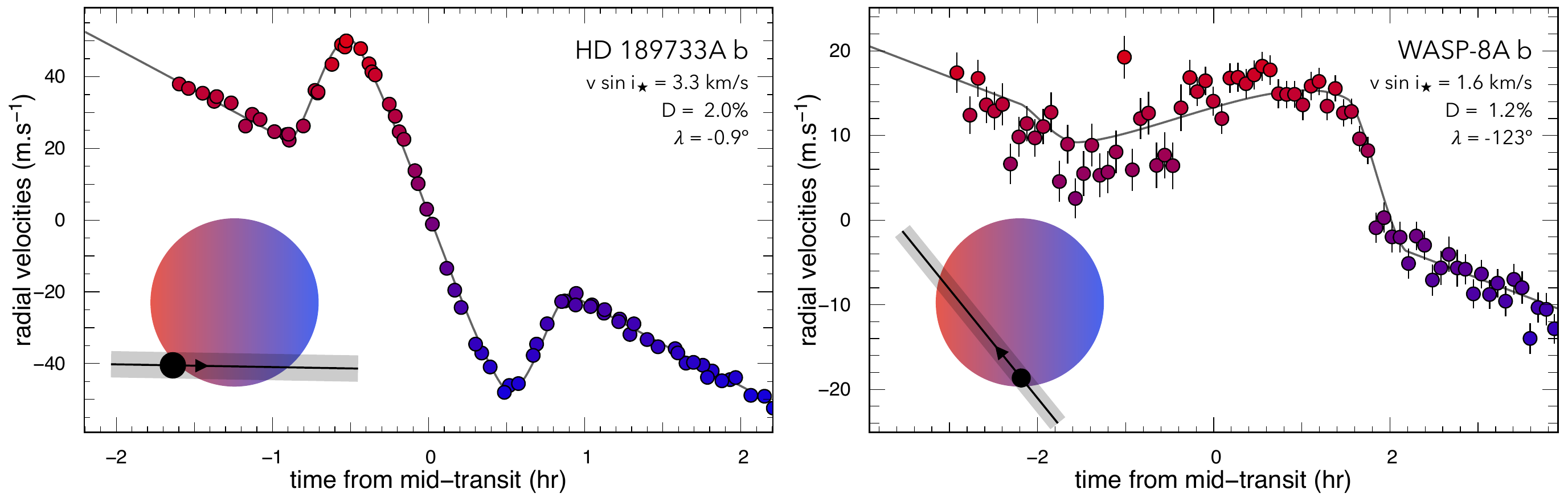}
\caption{The {Rossiter--McLaughlin effect} obtained on two transiting {hot Jupiter}s, with the HARPS spectrograph. The left panel shows HD\,189733A\,b, and on the right is plotted WASP-8A\,b. The radial-velocity measurements are colour-coded as a function of their Doppler information. The radial-velocity slope corresponds to the Doppler reflex motion of the host star, as caused by the orbiting planet. A most likely model is adjusted to the data, calculated from \citet{Gimenez:2006yq,Gimenez:2006kx}. Those measurements are reproduced from the analyses presented in \citet{Triaud:2009qy} and \citet{Queloz:2010lr} respectively. Visual representation of the orbits is provided on the bottom left of each panel. }
\label{fig:ros}       
\end{figure*}

The {Rossiter--McLaughlin effect} is the detection of a planetary {transit} using spectroscopy. It appears as an anomalous radial-velocity variation happening over the Doppler reflex motion that an orbiting planet imparts on its rotating host star (Fig.~\ref{fig:ros}). The shape of the {Rossiter--McLaughlin effect} contains information about the ratio of the sizes between the planet and its host star, the rotational speed of the star, the impact parameter and the angle $\lambda$ (historically called $\beta$, where $\beta = -\lambda$), which is the sky-projected {spin--orbit angle}.

The {Rossiter--McLaughlin effect} was first reported for an exoplanet, in the case of HD\,209458\,b, by \citet{Queloz:2000rt}. This effect takes its name from a pair of papers published jointly by \citet{Rossiter:1924qy} and \citet{McLaughlin:1924uq}, although earlier mentions exist. \citet{Schlesinger:1910fg} interprets discrepant radial-velocities obtained during an eclipse correctly, and \citet{Holt:1893fk} theorised about it much earlier than 1924. Currently the most complete collection of planetary {spin--orbit angle}s can be found on \href{http://www.astro.keele.ac.uk/jkt/tepcat/}{TEPCAT}\footnote{\href{http://www.astro.keele.ac.uk/jkt/tepcat/}{www.astro.keele.ac.uk/jkt/tepcat/}} \citep{Southworth:2011kx}. To date, the catalogue lists 181 measurements/analyses made on 109 planets. 

Since \citet{Queloz:2000rt}, measuring the {spin--orbit angle} has become a staple of the orbital characterisation of planetary systems \citep[e.g.][]{Bouchy:2005lr}. Observations of the {Rossiter--McLaughlin effect} are now obtained routinely and provide observables able to inform a debate on the various migratory histories of exoplanets. The most astonishing result was the discovery that like for many other observables, exoplanets present a large diversity in their {spin--orbit angle} \citep[e.g.][]{Hebrard:2008mz,Moutou:2009qv,Johnson:2009fv,Schlaufman:2010fk,Lendl:2014yu,Anderson:2015lr}. Nearly a third of known {hot Jupiter}s occupy orbits that are significantly misaligned with their host star \citep[see Fig.~\ref{fig:full}, and][]{Triaud:2010fr,Winn:2015lr}, with many on retrograde orbits \citep[e.g.][]{Winn:2009lr,Narita:2009lr,Anderson:2010fj,Queloz:2010lr,Collier-Cameron:2010lr,Addison:2013rm,Esposito:2014lr}. 

The distribution of $\lambda$ (or rather, of $|\lambda|$) with other (planetary or stellar) parameters can also be used to study the exchange of angular momentum between the planet and its host star, via tidal forces \citep[e.g.][]{Winn:2010rr,Triaud:2011fk,Guillochon:2011fk,Hebrard:2011fk,Brown:2011lr,Albrecht:2012lp,Dawson:2014kx,Petrovich:2015yu,Anderson:2015lr,Winn:2015lr}.

Our understanding of the {Rossiter--McLaughlin effect} has matured to a point where it can be used to confirm that an object is a planet, when RV precision is sufficient to rule out companion stars and brown dwarf, but too poor to enable the measure of a planetary mass \citep[e.g.][]{Collier-Cameron:2010lr,Zhou:2016rr,Temple:2017zl,Gaudi:2017eu}.

The {Rossiter--McLaughlin effect} can also be used to study the orbital inclination of eclipsing binary stars (its first application), and could be used to measure the rotational spin direction of exoplanets. It can provide information about an exoplanet's atmospheric composition as well as be used in the context of exo-moons. Further descriptions of these can be found towards the back of the chapter. Before closing this chapter, we will also describe a few alternate methods that have been devised to estimate the projected {spin--orbit angle}, without requiring an observation of the {Rossiter--McLaughlin effect}. A few other techniques also provide related observables.

\section{Basic concept}

A rotating star has one of its hemispheres blue-shifted and its other hemisphere red-shifted. This happens because one of its sides is approaching the observer while the other recedes. While the planet transits, it will cover different sections of the star sequentially. For instance, it can start to cover the blue-shifted side. In this case the average flux received by the observer will appear offset redward, creating a positive shift in the radial-velocity measured for the star, deviating from the Doppler reflex motion. As the planet scans the stellar velocity profile, the star appears to experience a rapid change in velocity. This is the {Rossiter--McLaughlin effect} (Fig.~\ref{fig:ros}). A detailed introduction can be found in \citet{Gaudi:2007vn}.

The shape of the {Rossiter--McLaughlin effect} contains several pieces of information. The most sought-after is that we can measure how long the planet spends over one hemisphere versus the other, which tells us the projected angle between the planet's path and the stellar equator. As can be expected, to a first order, the semi-amplitude of the {Rossiter--McLaughlin effect} scales with the planet's size and the stellar rotational velocity:
\begin{equation}\label{eq:firstorder}
A_{\rm RM} \simeq {2 \over 3}\,D\,v\sin i_\star\,\sqrt{1-b^2}
\end{equation}
where  $D = (R_{\rm p} / R_\star)^2$ is the {transit} depth, relating $R_{\rm p}$ (the planet's radius) to $R_\star$ (the host star's radius), where $v$ is the rotational velocity of the star (at the equator), $\sin i_\star$ is the inclination on the sky of the stellar rotation axis, and $b$ is the impact parameter (in units of $R_\star$). For a typical {hot Jupiter} with a {transit} depth of order 1.5\%, and a typical star with a rotational velocity of 2~km~s$^{-1}$, we obtain a semi-amplitude of  order  20~m~s$^{-1}$. The duration of the event is as long as the {transit}, which for a {hot Jupiter} is often about two to three hours. The Rossiter--McLaughlin signal can be detected when collecting high-cadence radial-velocity measurements made with high-precision, stable, high-resolution spectrographs such as those routinely used in the search for exoplanets (see Fig.~\ref{fig:ros}). Because the ingress and egress have a typical timescale of 20-30 minutes, a maximum recommended exposure for each spectrum is 10-15 minutes, but this can be adapted depending on the stellar and planetary parameters.

\section{Modelling the Rossiter--McLaughlin effect}


\subsection{The classical {Rossiter--McLaughlin effect}}

The earliest attempt to model the {Rossiter--McLaughlin effect} were made by \citet{Petrie:1938yg}. At that time, the {Rossiter--McLaughlin effect} was referred to as the {\it rotation effect} (or {\it rotational effect}). The principal formulations were then made by \citet{Kopal:1942ai}, but were only valid for coplanar systems. These calculations were generalised by \citet{Hosokawa:1953kl}, who introduced the sky-projected {spin--orbit angle}\footnote{for simplicity the sky-projected {spin--orbit angle} will often be referred to simply as the {spin--orbit angle}, within this document, something which is also done by several authors in the literature.} and called it $\beta$. Following this, a complete description was compiled by \citet{Kopal:1959pd,Kopal:1979uy}. Kopal used $\alpha$-{\it functions}, based on Legendre polynomials to integrate over the visible surface of the stars. They can account for multiple distortions of the shape, and of the luminosity distribution. While flexible, this approach is not the easiest to follow, were computationally intensive (in the absence of modern computers). Simpler formulations are now available. In models elaborated for transiting exoplanets, only one uses Kopal's formalism: \citet{Gimenez:2006yq,Gimenez:2006kx}.

Model adjustment on the first detection of the {Rossiter--McLaughlin effect} was made by drawing a grid over the star and reconstructing the cross correlation function of the stellar spectrum from which the radial-velocity is extracted \citep{Queloz:2000rt}. A grid based approach is being used by a few authors \citep[e.g.][]{Wittenmyer:2005bh,Covino:2013rf,Esposito:2017kx}. This can be handy in some situations, notably when having to also treat various stellar effects, such as stellar variability due to spot-induced rotational modulation, or the overlap of spots by a transiting planet \citep{Oshagh:2013fv,Oshagh:2016ys}.

The notation most frequently used for the {spin--orbit angle} is $\lambda$, which was introduced in \citet{Ohta:2005uq}\footnote{with $\lambda = -\beta$}. This presents an analytical expression for the {Rossiter--McLaughlin effect} that has been widely used. Improvement on this formalism were made first by \citet{Hirano:2011fk}, and by \citet{Boue:2013fk}, to resolve various issues related to different procedures used to extract radial velocities from stellar spectra. This can  affect the amplitude and shape the radial-velocity anomaly that is the {Rossiter--McLaughlin effect} \citep{Triaud:2009qy}. 
Please refer to \citet{Boue:2013fk} for a more ample description. Recently a study compared several methods of analysis, and confirmed that the Bou\'e model performs most consistently for instruments such as HARPS \citep{Brown:2017ul}. Additional corrections can be made, for instance to account for a small bias in the {spin--orbit angle} created by ignoring stellar convective blue-shift \citep{Shporer:2011fk}. We caution here that not all stellar effects are accounted for in the modelling of the {Rossiter--McLaughlin effect} \citep{Cegla:2016fr,Oshagh:2016ys,Reiners:2016lr}.

For references on models of planetary transits, including the {Rossiter--McLaughlin effect}, please be directed to the chapter by Deeg, H., ``Tools for Transit and Radial Velocity Modelling and Analysis'' for additional information.

\subsection{Some known pitfalls in model fitting}

Here are a few words of caution when attempting to adjust a model through the waveform of the {Rossiter--McLaughlin effect}. 
\begin{figure*}
\includegraphics[width=0.6\textwidth]{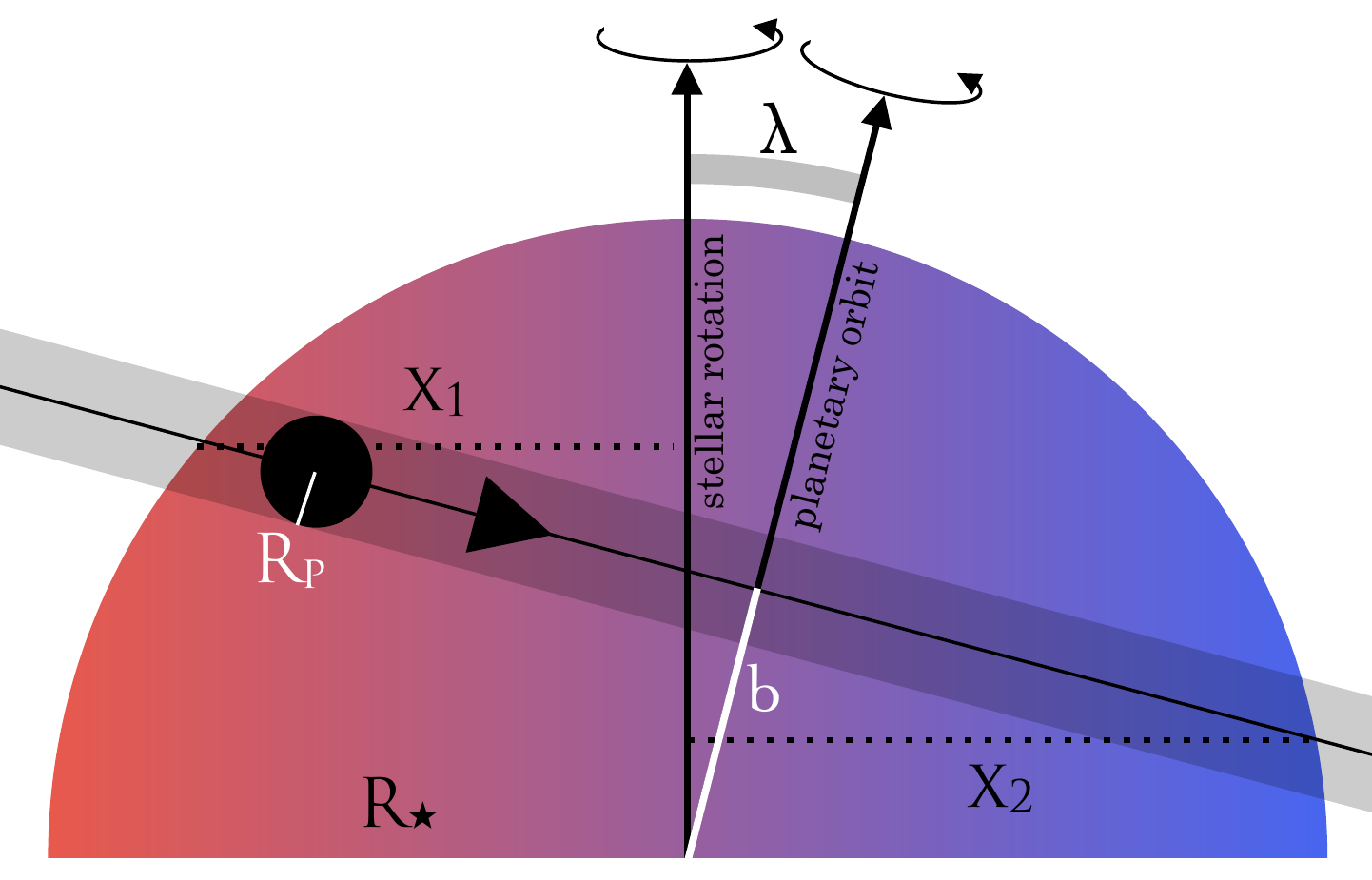}
\caption{Various quantities used to model the {Rossiter--McLaughlin effect}. Adapted from Fig. 1 in \citet{Albrecht:2011rt}.}
\label{fig:ros_b}       
\end{figure*}

\runinhead{When the impact parameter vanishes.}

As the impact parameter $b$ tends to zero, a degeneracy occurs between $v \sin i_\star$ and $\lambda$. This happens because in this instance, a heavily misaligned planet would produce a symmetrical effect: No matter how inclined the orbit is, the planet spends as much time on both hemispheres. A small $\lambda$, and low $v \sin i_\star$, produce an equivalent effect to a high $\lambda$, high $v \sin i_\star$ when $b=0$. This can be partially resolved if the stellar rotation is known and used as a prior \citep[e.g. WASP-1\,b,][]{Albrecht:2011rt}. However the stellar rotation as measured from spectral line broadening, and the values obtained from the  {Rossiter--McLaughlin effect} are not always compatible \citep[e.g.][]{Triaud:2011vn,Triaud:2015fk,Brown:2017ul}. An improper use of priors can also lead to spurious results as will be shown in the next paragraph.

What happens when $b = 0$ has been detailed in \citet{Albrecht:2011rt} and is here reproduced. Let's define two distances, $x_1$ and $x_2$ (as show in Fig.~\ref{fig:ros_b}), that can be shown to equate to:
\begin{equation}\label{eq:x1}
\begin{split}
x_1 &= \sqrt{1-b^2} \cos \lambda - b \sin \lambda \\
x_2 &= \sqrt{1-b^2} \cos \lambda + b \sin \lambda
\end{split}
\end{equation}
Following this we calculate the scaled sum of $x_1$ and $x_2$:
\begin{equation}\label{eq:x+y}
{1 \over 2} v \sin i_\star (x_1 + x_2) = \sqrt{1-b^2} v \sin i_\star \cos \lambda,
\end{equation}
and the scaled difference of  $x_1$ and $x_2$:
\begin{equation}\label{eq:x-y}
{1 \over 2} v \sin i_\star (x_1 - x_2) = b v \sin i_\star \sin \lambda
\end{equation}

Eq.~\ref{eq:x+y} measures the semi-amplitude, and eq.~\ref{eq:x-y} measures the asymmetry between both extrema. In eq.~\ref{eq:x-y}, we can see that when $b = 0$, the difference vanishes and we can only properly estimate the semi-amplitude, which is a function of $v \sin i_\star$ and $\cos \lambda$. For an extended discussion please refer to \citet{Albrecht:2011rt}.

Because of this degeneracy, applying priors on $v \sin i_\star$ can lead to astonishing results: WASP-23\,b is example with $b \sim 0$. Two different estimates of the stellar rotation were computed. Of both values, the faster rotation leads to a conclusion where WASP-23\,b occupies a nearly perpendicular orbit to its host's equator, whereas the slower measurement of rotation makes $\lambda$ compatible with an aligned solution \citep{Triaud:2011vn}.

\runinhead{Spin--orbit angles and non-detected effects.}

If the star rotates, and if the planet transits, then there must be a {Rossiter--McLaughlin effect} in the spectroscopic data. However the effect is not always detected. Sometimes this is because of measurement uncertainties. At other times, the orbital configuration weakens the amplitude of the effect (e.g a situation close to  $b=0$, $\lambda=90^\circ$). In both instances, good care must be taken. 

A numerical experiment realised in \citet{Albrecht:2011rt} demonstrated that fitting for a {Rossiter--McLaughlin effect} through randomly distributed radial-velocity measurements (in the absence of a {Rossiter--McLaughlin effect}), leads to favoured angles around $0^\circ$ and $180^\circ$. This is the main cause behind a faulty conclusion on WASP-2\,b \citep{Triaud:2010fr}. Initially the data was interpreted as showing evidence of a retrograde planet, but \citet{Albrecht:2011rt}  convincingly showed that instead the event is not detected.

\citet{Albrecht:2011rt} also analysed WASP-1\,b, and do not detect of the {Rossiter--McLaughlin effect}, but can conclude about its misalignment thanks to the high rotation rate of the star. Two similar cases with different outcomes.

\runinhead{Combining parameters.} 
Usually, we know a planet transits before attempting Rossiter-McLaughlin observations. We already have precisely measured most of the {transit} parameters. Fitting the {Rossiter--McLaughlin effect} essentially entails the adjustment of only two additional variables: $v \sin i_\star$ and $\lambda$. As $v \sin i_\star$ gets close to zero, or if $v \sin i_\star$ has a large uncertainty, $\lambda$ becomes ill defined. Those two parameters are highly correlated in a non-linear way (as touched upon just above). This means that exploring parameter space can become inefficient when using Markov Chain Monte Carlo (MCMC) algorithms and can trick various optimisers into local minima. Furthermore, since a negative $v \sin i_\star$ is unphysical (although it could be interpreted as a retrograde planet), the value is often forced to remain positive. This can overestimate the detection significance of the {Rossiter--McLaughlin effect}.

If we examine eq.~\ref{eq:x+y}~\&~\ref{eq:x-y}, we can notice that $v \sin i_\star$ and $\lambda$ can be reparametrised into $(v \sin i_\star \cos \lambda)$ and $(v \sin i_\star \sin \lambda)$. This is analogous to the orbital eccentricity $e$, which can be combined to its angle of periastron $\omega$ into ($e \cos \omega$, $e \sin \omega$) \citep{Ford:2005qy}. However, as noted in \citet{Ford:2006yq}, when sampling uniformly, this creates a prior that increases linearly with $e$ instead of being un-informative. This biases $e$ towards higher values, and led to a series of spurious detections of eccentricities for short-period {hot Jupiter}s. 

This also happened during the original analysis of WASP-2\,b \citep[in][]{Triaud:2010fr} and called upon by \citet{Albrecht:2011rt}. In the case of eccentricity the problem was remedied by using instead ($\sqrt{e} \cos \omega$, $\sqrt{e} \sin \omega$). A similar operation can be made for our parameters of interest: ($\sqrt{v \sin i_\star} \cos \lambda$, $\sqrt{v \sin i_\star} \sin \lambda$) \citep{Triaud:2011vn}.

\subsection{Alternate ways of modelling the Rossiter--McLaughlin effect}

\runinhead{Doppler tomography.} The traditional way of modelling the {Rossiter--McLaughlin effect} is via a radial-velocity timeseries. The information contained in the radial-velocity originates from mis-shaped stellar absorption lines, caused by a deficit of flux on the blue or red-shifted hemisphere. It is challenging to detect temporal variations in the shape of a single line by a transiting planet, although it has been achieved for some eclipsing binaries \citep{Albrecht:2007th,Albrecht:2009fy}. In the situation of a planet, this is solved by collecting multiple lines (and thousands exist in the wavelength range of most high-resolution spectrographs), sometimes in the form of a cross-correlation function. The study of a line shape is called {Doppler tomography} and has been used in a variety of applications, like mapping inhomogeneities on stellar surfaces, investigate magnetic fields, study binaries, and their accretion flows, but also planetary atmospheres, resolve planetary orbital motion, as well as reveal cloud patches in brown dwarf atmospheres \citep[e.g.][]{Marsh:1988ul, Richards:1995pd, Collier-Cameron:1998ly,Donati:2006fj, Snellen:2010lr, Rodler:2012qy, Brogi:2012fk, Crossfield:2014db,Richards:2014gf}.

The mean, out-of-transit, line shape is removed from all epochs. If a dark inhomogeneity exists on the stellar surface, it appears as a positive bump, that will appear at ingress, travel from one side of the line to the other side and disappear at egress. We call this signal the planet's {\it trace}, or the planet's {\it Doppler shadow}. This was first done on HD\,189733A\,b by \citet{Collier-Cameron:2010lk}, and the technique was quickly adopted to confirm WASP-33\,b \citep{Collier-Cameron:2010lr}. In this particular case the planet orbits a pulsating $\delta$ Scuti, in a retrograde configuration. WASP-33\,b's Doppler shadow offers an angle, to the signal that stellar pulsations produce in the tomogram, and which rotate in and out of view in the direction of rotation.

{ Doppler tomography} has now been adopted by several authors \citep[e.g.][]{Brown:2012fk,Gandolfi:2012lr,Albrecht:2013lr,Zhou:2016dq,Temple:2017zl} and is most often used for faster rotating stars where the Doppler shadow is easier to resolve, and where radial-velocity uncertainties increase. The Doppler shadow is modelled with similar observables to those used for the {Rossiter--McLaughlin effect}. The planet's bump also carries information about the stellar rotation. It was expected that $v \sin i_\star$ would be more accurately determined and that the model adjustment to the data would be less sensitive to degeneracies arising when the impact parameter approaches zero. However a comparison analysis did not show a clear preference for this approach \citep{Brown:2017ul}. Of interest is that the planet trace forms a signal close to a line, which is a simpler shape to adjust a model to than the usual succession of three slopes seen on Fig.~\ref{fig:ros}. Tomography can in principle be more forgiving about the observing cadence than the traditional {Rossiter--McLaughlin effect} would. Contrary to the {Rossiter--McLaughlin effect}, the Doppler shadow does not become 0 when the planet crosses the stellar axis, which can also help boost the significance of a detection and study configurations near $\lambda = 90^\circ$ \citep[e.g. WASP-76\,b;][]{Brown:2017ul}.

\runinhead{The Rossiter--McLaughlin effect reloaded.}

An additional method was recently devised. It relies on reconstructing the velocity field that the planet hides during {transit}, using line profiles (or cross correlation functions). A description can be found in \citet{Cegla:2016qy}, where this method is applied on HD\,189733A\,b. This new approach is ambitious and can recover the latitudinal {differential rotation}, the true axis of the star $i_\star$, and account for convective blueshift and its centre to limb variation. Having measured $i_\star$ the authors could also compute $\psi$, the true {spin--orbit angle}, instead of $\lambda$, $\psi$'s sky-projection. There are also indications that spot-crossing events can be identified. A recent re-analysis of the WASP-8A\,b {transit} \citep{Bourrier:2017lq} yielded a {spin--orbit} significantly different to its original, traditional analysis \citep{Queloz:2010lr}, although the planet remains retrograde.

\section{Measuring $\psi$}

The traditional {Rossiter--McLaughlin effect} only provides information on sky-projected {spin--orbit angle}, $\lambda$. However, the true  {spin--orbit angle} $\psi$ is the quantity that is really sought after. Statistical comparison between theoretical $\psi$ distributions \citep[e.g.][]{Wu:2007ve,Fabrycky:2007pd,Chatterjee:2008uq,Nagasawa:2008gf,Naoz:2012fk,Petrovich:2015yu} and the observed $\lambda$ can be made. This is done either by projecting $\psi$ on the sky, or de-projecting $\lambda$ \citep[][]{Fabrycky:2009vn,Triaud:2010fr,Brown:2012lr}. \citet{Fabrycky:2009vn} relate $\psi$ to $\lambda$:
\begin{equation}
\cos \psi = \cos i_\star \cos i_{\rm p} + \sin i_\star \sin i_{\rm p} \cos \lambda
\end{equation}

However, the best correction would be no correction at all. The key in estimating $\psi$ lies in measuring $i_\star$ properly, which is a hard measurement to make. There exist  various ways but the most frequent is to transform a determination of the rotation period of the star (whose radius is often accurately known) into an equatorial velocity $v$, and solve for $\sin i_\star$. Complication arise because $v \sin i_\star$ as measured via fitting the {Rossiter--McLaughlin effect} can differ from the value usually estimated from spectral line broadening \citep{Triaud:2011vn,Triaud:2015fk}, and can be affected by latitudinal {differential rotation}. Moreover, determining the rotation period of a star can be fraught with uncertainties too. Stellar rotation is often measured by the photometric modulation caused by spots rotating in and out of view as the star rotates. Spots can be multiple, and located at different latitudes, providing different rotation periods. Empirical relations between activity levels and rotation period \cite[e.g.][]{Mamajek:2008lr} are also used, but their accuracy for single objects remain far from guaranteed. This has not prevented many authors to produce estimates of $\psi$ \citep[e.g.][]{Winn:2009lr,Hebrard:2011lr,Sanchis-Ojeda:2011zr,Hellier:2011lr,Lendl:2014yu,Anderson:2015lr,Zhou:2016rr,Esposito:2017kx}. These attempts are very informative, but should be taken more critically than $\lambda$.

The most robust determination of $\psi$ so far, was probably brought forward thanks to {asteroseismology}. Rotation induces a splitting of oscillation modes, that can be used to compute $i_\star$ \citep{Gizon:2003nr}. This has provided an alternative method to measuring the {spin--orbit angle} of a planetary orbit \citep[see later;][]{Chaplin:2013zr,Huber:2013jk,Van-Eylen:2014xy}, but combined with the {Rossiter--McLaughlin effect}, this leads to $\psi$ \citep[][]{Benomar:2014cr,Campante:2016wd}.

\section{Rossiter-McLaughlin measurements on hot Jupiters }

Apart from measurements on HAT-P-11\,b, a near-polar Neptune-sized object \citep[e.g.][]{Winn:2010lr,Hirano:2011ye}, an attempt on GJ\,436\,b, another Neptune \citep{Albrecht:2012lp}, and two attempts on 55\,Cnc\,e, a super-Earth \citep{Lopez-Morales:2014qv,Bourrier:2014lr}, the bulk of the observations have concentrated on gas giants, and particularly on {hot Jupiter}s. 

\runinhead{Setting the scene.} {Hot Jupiter}s are loosely defined. Typically, they are planets with masses in excess of 0.1~M$_{\rm Jup}$, and periods shorter than $\sim 10$ days of which 51\,Peg\,b is a good (and the first) example \citep{Mayor:1995uq} (for more information, please read the chapter by Santerne, A. ``Hot Jupiter Populations from Transit and RV Surveys''. The population is identified by an over-density around a three day orbital period, for a G dwarf host \cite[e.g.][]{Santerne:2016lr}. There is considerable debate on how {hot Jupiter}s are formed. \\

Hot Jupiters could theoretically assemble on orbits close to those we observe them on \citep{Bodenheimer:2000lr,Batygin:2015lr}. However the lack of nearby sub-Neptune companions to {hot Jupiter}s contradicts this scenario \citep{Huang:2016yq}. Only one {hot Jupiter} escapes this otherwise general rule: WASP-47\,b \citep{Becker:2015fk,Neveu-VanMalle:2016xy}. The situation is different for gas giants on orbits longer than 10 days, sometimes dubbed {\it warm Jupiters}. They exist frequently alongside close-by companions, which informs us that warm and {hot Jupiter}s constitute two separate populations of gas giants. Those two populations probably overlap, and it is credible that WASP-47\,b happens to be amongst the shortest orbiting examples of the warm population \citep{Huang:2016yq}.\\

\begin{figure*}
\includegraphics[width=\textwidth]{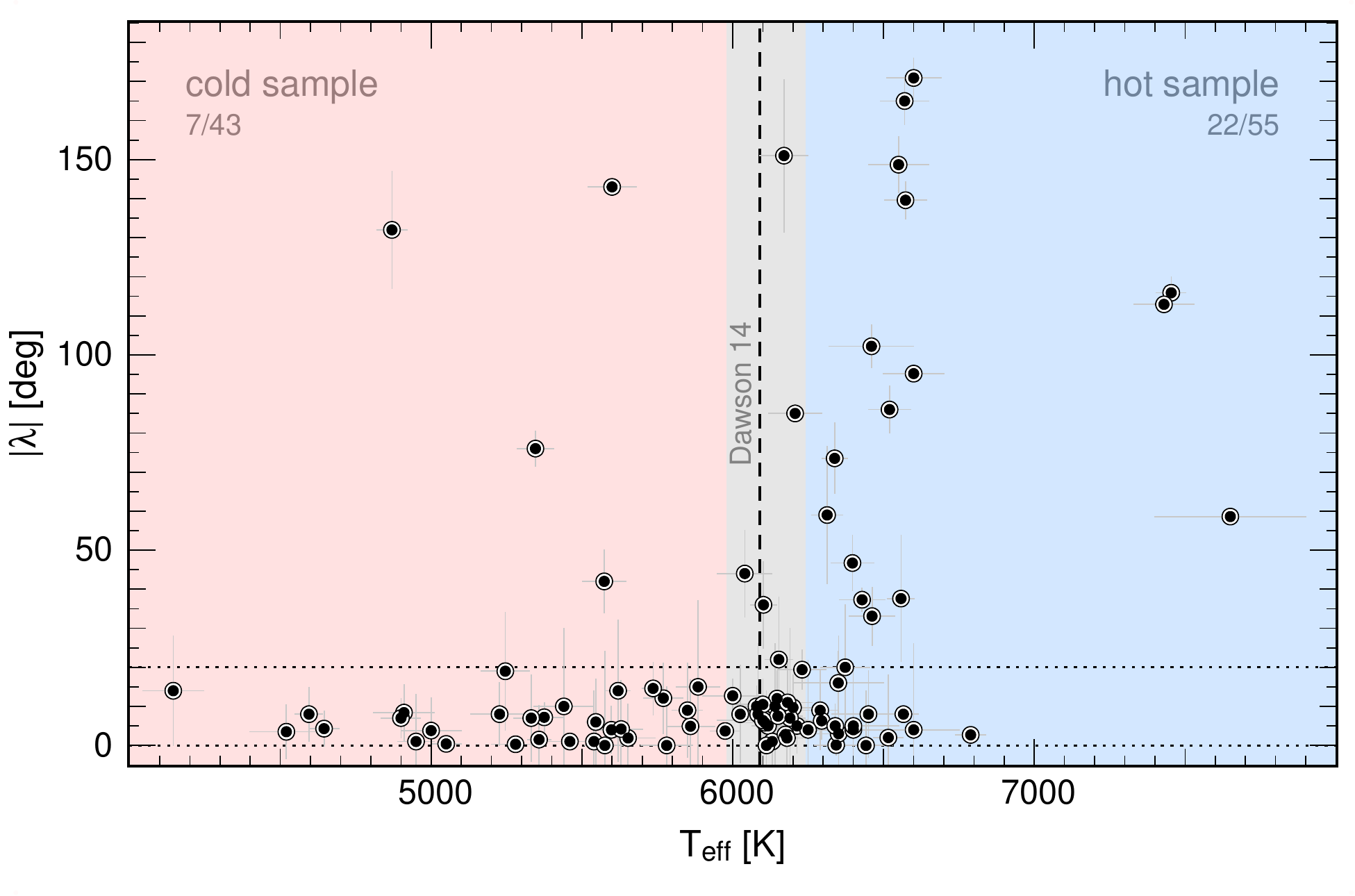}
\caption{Scatter plot showing the absolute values of the projected {spin--orbit angle}s $\lambda$ as a function of the host star's effective temperature, for gas giants. A cold and a hot sample are separated by a transition effective temperature as determined by \citet{Dawson:2014kx}. The most likely value (6090 K) is shown with a vertical dashed line, and the grey region around it is the $1\sigma$ confidence region. The two dotted horizontal lines contain planets that can be considered on coplanar orbits. The numbers show the fraction of planets that are non-coplanar on either side of the dashed line. KELT9\,b escapes this diagram on the right hand side (and is non-coplanar).}
\label{fig:full}       
\end{figure*}

\begin{figure*}
\includegraphics[width=\textwidth]{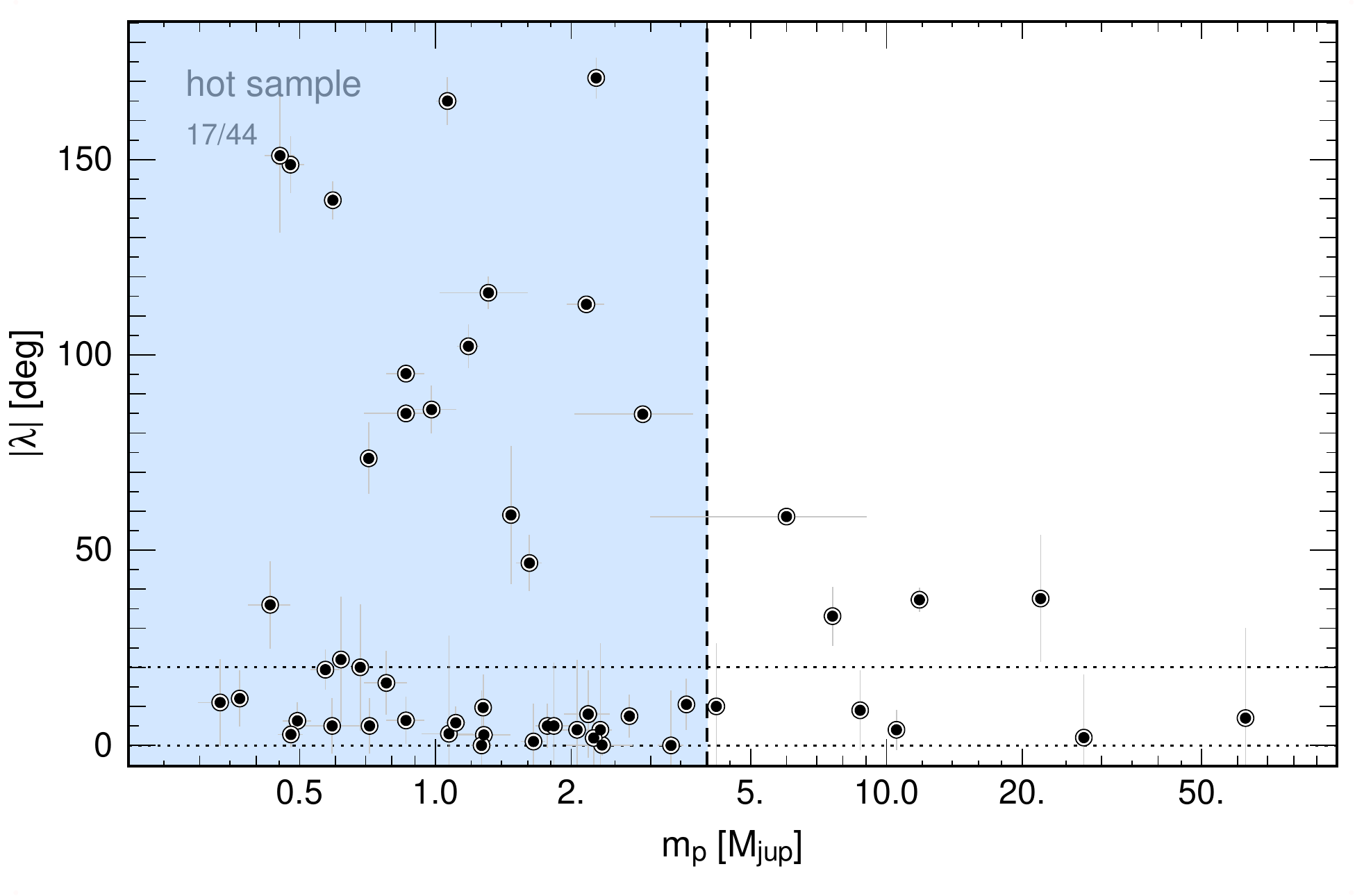}
\caption{Scatter plot showing the hot sample of $\lambda$ values, defined in Fig.~\ref{fig:full}, and ordered against the planet's mass. The two dotted horizontal lines contain planets that can be considered on coplanar orbits. A lack of retrograde planet was noticed by \citet{Hebrard:2011lr} and is visible here too, for masses $> 3 M_{\rm Jup}$ (vertical dashed line). The number shows the fraction of planets that are non coplanar on the left hand side of the dashed line. }
\label{fig:hot}       
\end{figure*}

\begin{figure*}
\includegraphics[width=\textwidth]{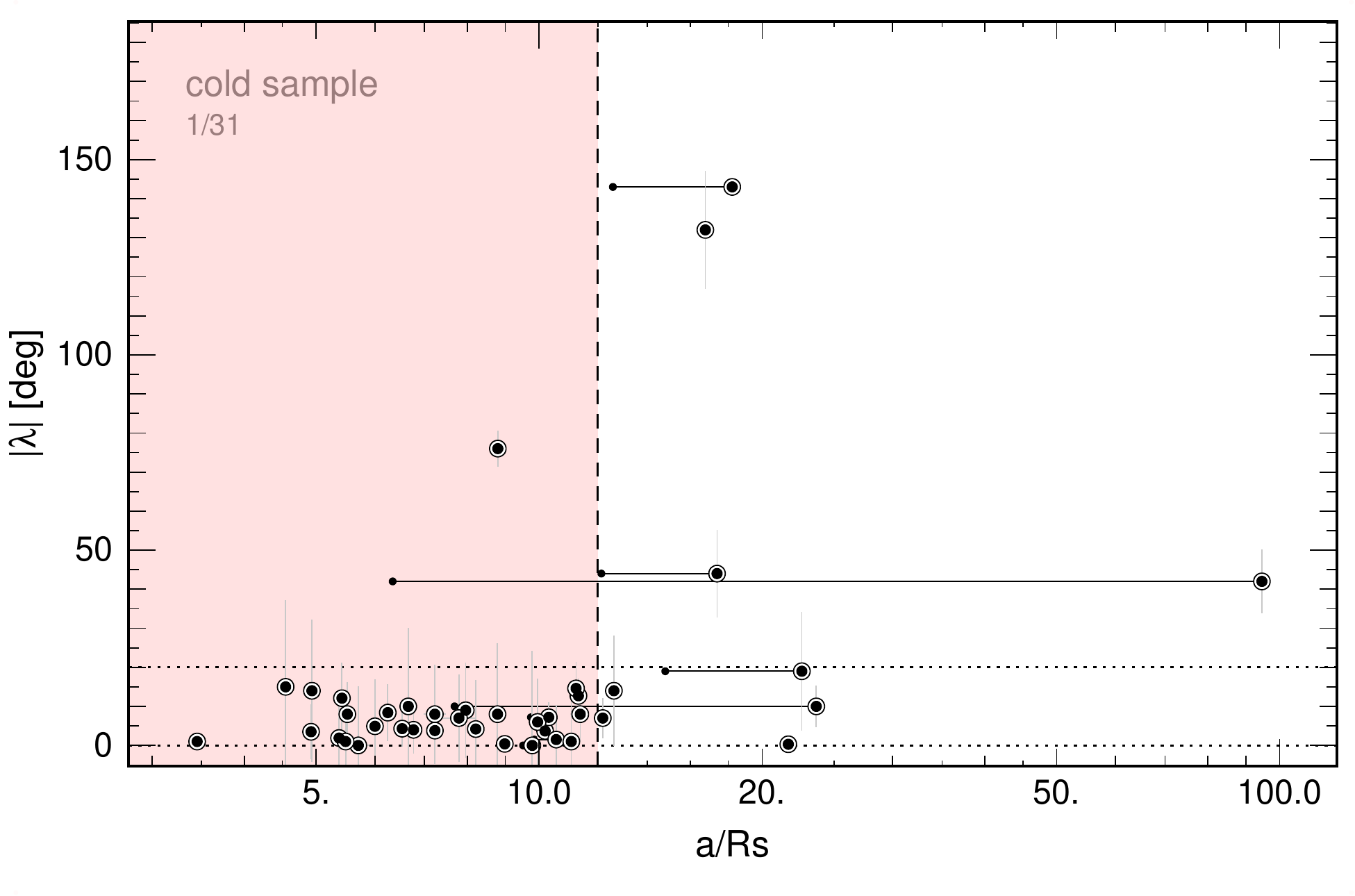}
\caption{Scatter plot showing the cold sample of $\lambda$ values, defined in Fig.~\ref{fig:full}, and ordered versus the orbital separation in units of stellar radii ($a/R_\star$). Orbital eccentricities are depicted with a thin horizontal line ending with a dot, at the distance of periastron. The two dotted horizontal lines contain planets that can be considered on coplanar orbits. 
A lack of misaligned and circular orbits is seen for $a/R_\star < 12$ (vertical dashed line) \citep{Anderson:2015lr}. The number shows the fraction of planets that are non coplanar on the left hand side of the vertical dashed line. }
\label{fig:cold}       
\end{figure*}

\begin{figure*}
\includegraphics[width=\textwidth]{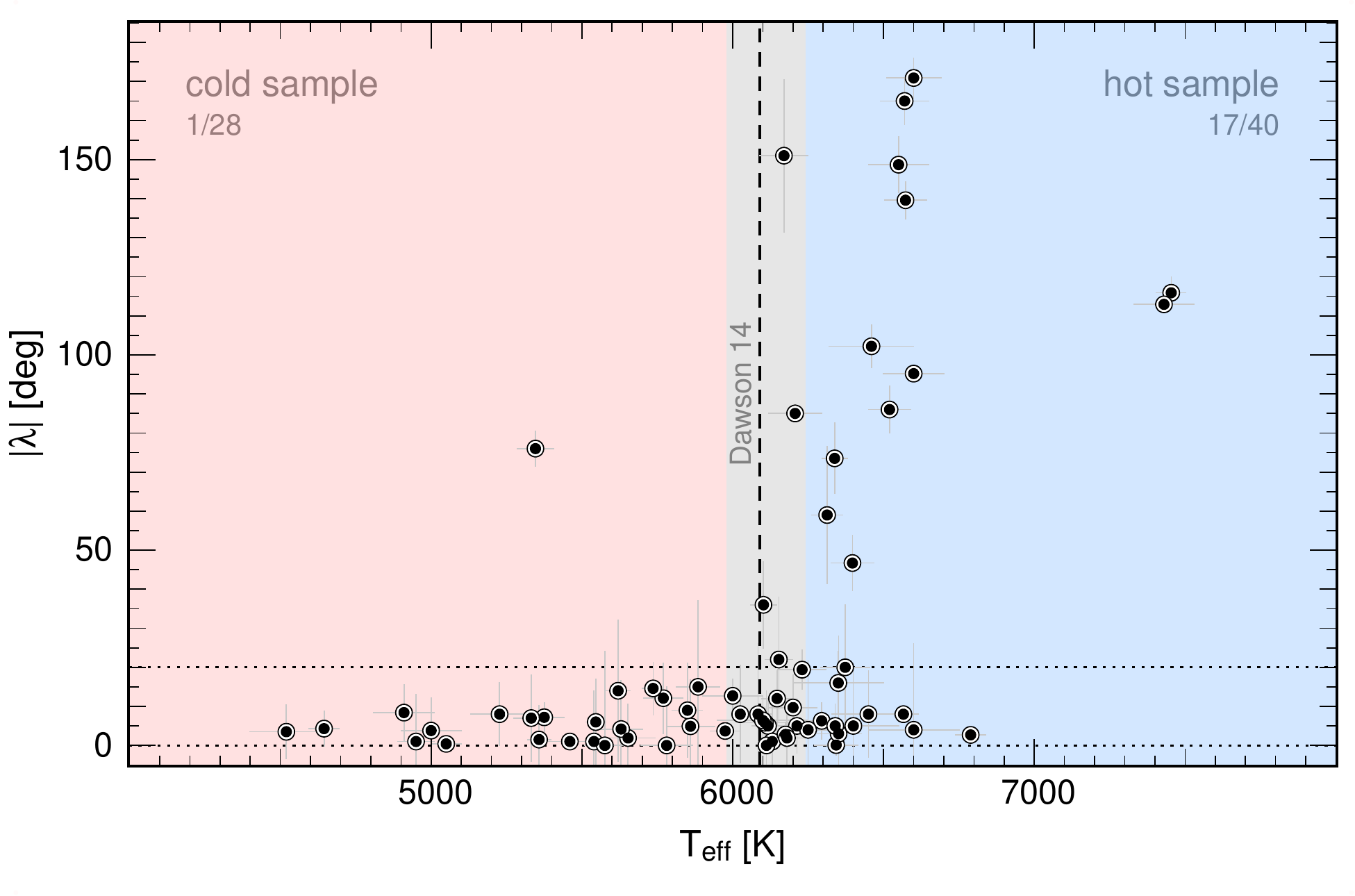}
\caption{Reproduction of Fig.~\ref{fig:full} after the removal of patterns shown in Fig.~\ref{fig:hot} \& \ref{fig:cold}. The cold and a hot sample are more contrasted than in Fig.~\ref{fig:full}. The two dotted horizontal lines contain planets that can be considered on coplanar orbits. The numbers show the fraction of planets that are non coplanar on either side of the dashed line.}
\label{fig:clean}       
\end{figure*}

The alternative is that {hot Jupiter}s form further from their star \citep[e.g.][]{Rafikov:2006lr}, and migrate inwards. This orbital migration takes two main flavours: disc-driven migration, and disc-free migration.

Disc-driven migration is expected to keep gas giants within the plane of the disc as the planet exchanges angular momentum with the disc material to reduce its semimajor axis \citep{Goldreich:1980lr,Lin:1996yq,Ward:1997kx,Baruteau:2014rp}. If the whole disc is misaligned \citep{Batygin:2012hl,Lai:2014nx} with the stellar equator \citep[which is not observed;][]{Greaves:2014vn}, or if the disc's outer parts are warped \citep{Terquem:2013qy}, gas giants can end up on inclined orbits. 

Disc-free migration contains a vast number of scenarios that come (so far) under one general umbrella: dynamical interactions placing a planet on an eccentric orbit, whose semimajor axis decays thanks to tidal dissipation when the planet is at periastron. This can be caused by other planets in the system \citep{Rasio:1996ly,Nagasawa:2008gf,Chatterjee:2008uq,Wu:2011ul,Naoz:2011lr,Petrovich:2015yu}, other stars \citep{Wu:2007ve,Fabrycky:2007pd,Malmberg:2011dq,Naoz:2012fk,Petrovich:2015qf} or a disc \citep{Terquem:2010fj,Terquem:2013qy}.\\

We could also imagine that the two processes can be mixed together, with planets getting closer to their star first by disc-driven accretion, which ceases but is followed by a disc-free, dynamical and tidal migration.

These multiple scenarios can produce both aligned and inclined planets making inferences hard to make. This is compounded by the fact that tides likely affect the distribution of {spin--orbit angle}s, and that sometimes the stellar axis of rotation itself may not be stable. It might vary with respect to a planetary orbit, during the disc phase \citep{Lai:2011ul}, or afterwards \citep{Cebron:2013dq,Rogers:2013la}.

For a more rigorous discussion on planetary migration, the reader should consult the chapter by Nelson, R., ``Planetary Migration in Protoplanetary Disks''.
 
\runinhead{The measurements.} Fig.~\ref{fig:full} to \ref{fig:clean} represent the population as catalogued in TEPCAT (2017-06-26). Only planets with masses $> 0.1 M_{\rm Jup}$ are represented. A number of measurements have also been curated out on account of uncertain parameters, degenerate solutions, or concerns with the analysis. {Spin--orbit angle}s $\lambda$ with uncertainties greater than $30^\circ$ are also discarded. This leaves 93 systems, 26 of which are misaligned, which are represented in Fig.~\ref{fig:full}. Those systems can be considered as coming from two populations, a {\it cold} and a {\it hot} sample, separated at around $T_{\rm eff} = 6090^{+150}_{-110} K$ \citep{Dawson:2014kx}. The cold sample is typically formed of coplanar planets, whereas the hot sample contains a much larger occurence of inclined orbits. This pattern was noticed by \citet{Schlaufman:2010fk} and \citet{Winn:2010rr}, and confirmed in \citet{Albrecht:2012lp} and \citet{Brown:2017ul}. A likely interpretation is provided by \citet{Dawson:2014kx} and involves tides, as well as the spin-down of stars thanks to magnetic braking.  \\

Scrutinising the cold and hot samples in isolation, they too come with patterns. Within the hot sample, we can observe that planets with masses above $\sim 3 M_{\rm Jup}$ are less likely to be retrograde \citep[Fig.~\ref{fig:hot};][]{Hebrard:2011lr}. This is understood to be because tidal realignment scales with $(M_\star/m_{\rm p})^2$ \citep{Zahn:1977yq,Barker:2009fk,Dawson:2014kx}.

The cold sample is interesting too, if ordered as a function of the semimajor axis in units of stellar radii ($a/R_\star$), which is a parameter directly determined from a {transit} lightcurve. For $a/R_\star > 10-15$, we observe a larger fraction of eccentric and of misaligned orbits. A first version of Fig.~\ref{fig:cold} is presented in \citet{Anderson:2015lr}. Here, the periastra are also represented with a horizontal line ending in a small dot showing that some of the planets with larger separations are likely on their way to become {hot Jupiter}s (like HD\,80606\,b, the furthest to the right). This pattern can emerge thanks to tides since the timescale of orbital realignment scales with $(a/R_\star)^6$ \citep{Zahn:1977yq,Barker:2009fk,Dawson:2014kx}. It also worth reflecting on the objects that are at large orbital separation, but are aligned. These gas giants are too far from their star for tides to be important in the evolution of their orbital parameters. WASP-84\,b for instance, is circular, and aligned, at $a/R_\star > 20$ \citep{Anderson:2015lr}. This means it is consistent with having formed locally in a coplanar protoplanetary disc, or that it followed a disc-driven migration.\\

Now that two additional patterns have been noticed in the hot and cold sample, we can do a final exercise and remove all planets with masses $> 3 M_{\rm Jup}$ and $a/R_\star > 12$, to obtained a {\it cleaned} version of Fig.~\ref{fig:full} that is presented in Fig.~\ref{fig:clean}. The contrast between the cold and hot sample is now reinforced (a similar route is taken in \citet{Albrecht:2012lp}). The only misaligned object orbiting a cold star is HATS-14\,b \citep{Zhou:2015rz}.

\runinhead{Lessons from the data.}
More planets are coplanar than misaligned, and coplanarity becomes the norm for planets orbiting stars with $T_{\rm eff}< 6100$ on close, circular orbits. Patterns with planet mass, and planet separation are consistent with tidal realignment, but may also show different processes. For instance for $a/R_\star > 12$ we mostly have warm Jupiters instead of {hot Jupiter}s, and both populations are different \citep{Huang:2016yq}. If their formation, and/or orbital evolution are distinct, then the distribution in $\lambda$ can be expected to be distinct too.

In order to learn about how {hot Jupiter}s form we have to look in different ways. For instance, a different environment may provide different outcomes favouring one scenario over another. In open clusters, {hot Jupiter}s appear more frequent \citep[e.g.][]{Brucalassi:2016lr}, which can be interpreted as a dynamical origin \citep[e.g.][]{Triaud:2016fk}. Alternatively, the chemical make-up of transiting {hot Jupiter}s can be retrieved thanks to transmission and emission spectroscopy \citep{Deming:2017lr} (see Chapter by Kreidberg, L., ``Exoplanet Atmosphere Observations from Transmission Spectroscopy and Other Planet-Star Combined Light Observational Techniques'') . Different elemental abundances can be expected depending on the location of formation and the type of orbital migration \citep{Oberg:2011qv,Madhusudhan:2011hc,Madhusudhan:2017gf}. This would also provide one additional way of comparing the hot to the warm Jupiters, but also the aligned to the misaligned planets \citep{Huang:2016yq}.

\section{Alternate means to measure the spin--orbit angle of exoplanets}\label{sec:alt}

The {Rossiter--McLaughlin effect} and its derivatives are not the only means to measure an angle between a planet's orbital plane and it host star's equatorial plane. The success of  {Rossiter--McLaughlin effect} observations and the surprising results it produced, catalysed a palette of other techniques able to extract the {spin--orbit angle}, or a related observable. This proves convenient to explore areas of parameter space hard to reach with the traditional {Rossiter--McLaughlin effect}, or for stars that were too faint to conduct what remain fairly demanding and specialised observations.

\runinhead{Using photometry.}

Several methods were developed opportunistically, exploiting the availability of large, nearly un-interrupted photometric timeseries, of high quality acquired by the {\it Kepler} spacecraft. For instance, the shape of a planetary {transit} is controlled by the distribution of light covered by the passing planet. Fast rotating stars, instead of remaining spherical, become elliptical. The temperature is colder at the equator compared to the poles, with obvious effets on luminosity. This process is called {gravity darkening}, and distorts the standard {transit} shape \citep[e.g.][]{Barnes:2009qy}, particularly if the orbit is misaligned with its star. This is how it was deduced that KOI-13.01 (now Kepler-13A\,b) occupies an inclined orbit \citep{Szabo:2011fr,Barnes:2011uq}. A recent analysis computes its true obliquity $\psi$ by combining a tomographic measurement \citep{Johnson:2014db} with {gravity darkening} \citep{Masuda:2015uq}.

{\it Kepler} also permitted a boom in {asteroseismology}. It provides a very elegant and robust way of measuring the inclination of the stellar axis with respect to sky. Oscillations in a star have typical frequencies notably dependent on their internal density. These frequencies are isolated by constructing a power spectrum of the photometric timeseries, where clear modes emerge from the noise  \citep{Chaplin:2013lr}. Fast rotation can split some of those modes. This has been realised for a transiting planet host for the first time by \citet{Chaplin:2013zr}, and led to the stunning result of two super-Earths coplanar with one another, but whose common plane is misaligned with the equator of their host, Kepler-56 \citep{Huber:2013jk}.

Many Sun-like stars have spots. Spots create quasi-periodic modulations on a photometric timeseries, as they appear and disappear from the visible hemisphere, and as they evolve in size and latitude \citep[e.g.][]{Alonso:2009kl}. When a planet occults a spot, it also leaves a detectable imprint on its lightcurve \citep[e.g.][]{Pont:2007hl}. Particularly for stars rotating significantly slower than the planet's orbital period, the spot will have moved a little by the time the planet returns in {transit}. An inclined planet will more easily miss the spot as it rotates out of the {transit} chord, but an aligned planet is likely to encounter the spot on a few occasions \citep{Sanchis-Ojeda:2011xy,Sanchis-Ojeda:2011zr,Nutzman:2011db,mancini:2014zz,Mocnik:2016zr,Dai:2017cr}. The timing of starspot crossings can also be combined with out-of-transit variations to estimate the {spin--orbit angle} \citep{Sanchis-Ojeda:2012ys}. This is particularly effective when few spot crossing events are recorded, but does require a nearly continuous lightcurve.

\runinhead{Resolving the orbital inclination of non-transiting planets.}

\citet{Sahlmann:2011fk} determine the orbital inclination of a planet on the sky using {\it Hipparcos} astrometry, boosting its signature thanks to radial-velocity data. This inclination appears inconsistent with the inclination of the stellar axis, measured by comparing the stellar $v \sin i_\star$, and an activity indicator called $\log R'_{\rm HK}$, which is correlated to stellar rotation periods \citep[e.g.][]{Mamajek:2008lr}. This indicates the planet is likely misaligned.

When {\it GAIA} (Sozzetti, A., Bruijne, J., ``Space Missions for Exoplanet Science with Gaia and the Legacy of HIPPARCOS'' ) makes its final data release, we ought to receive several thousand planetary orbits \citep[e.g.][]{Perryman:2014rf} where this procedure can be reproduced. Those detections are expected to be gas giants, primarily, on orbital separation of a few AUs. They will provide a particularly interesting sample, in parts because we can compare it to the {hot Jupiter}s. Some of these {\it GAIA} detections will happen in systems where close-in planets are (or will become) known to {transit}, while remaining undetected astrometrically. Then we can be in a position to collect information on {mutual inclination}s \citep{Triaud:2017rm} and therefore on the {spin--orbit} alignment of distant planets. Similarly, it will become possible to measure the inclination between the orbital plane of a circumbinary planet and the plane of its central binary \citep{Sahlmann:2015xy}

Using CRIRES, \citet{Rodler:2012qy} and \cite{Brogi:2012fk} resolved the planet $\tau$\,Bo\"otis\,b spectroscopically, and could measure it orbital inclination on the sky and therefore, its true mass. Adopting this inclination for the star, its $v \sin i_\star$ yields a rotation period consistent with the planet's orbital period, indicative of having achieved tidal synchronisation, as was proposed by \citet{Donati:2008rm} based on Zeeman Doppler Imaging data. This strongly suggests that $\tau$\,Boo\,b is coplanar with its star, in line with the general behaviour of massive planets \citep{Hebrard:2011fk}.

\runinhead{Dynamical effects.}

The orbital inclination of circumbinary planets such as Kepler-16\,b \citep{Doyle:2011vn} is accurately known from the photometry, as the planet {transit}s first one star and then the other, and does so again at every orbital period. However, if the planet is but a little inclined with respect to the binary orbital plane, it will torque the planetary orbit causing it to precess or nutate \citep[e.g.][]{Doolin:2011lr}. This will move the planet in and out of {\it transitability} \citep{Schneider:1994lr,Martin:2014lr}. The most inclined circumbinary planet is Kepler-413\,b \citep{Kostov:2014qy}, with an inclination of $2.5^\circ$. 

On a similar note, a star rendered elliptical by its fast rotation will exert a torque a planetary orbit, changing its {transit} chord in a noticeable manner as described in \citet{Szabo:2011fr}, \citet{Barnes:2011uq}, \citet{Johnson:2015rr} and \citet{Masuda:2015uq}. The period precession encodes information on the inclination of the planet with respect to the star. 

\runinhead{Statistically.}

Most of these alternate methods work with observations obtained on singular systems. However it is also possible to approach the problem by examining the statistical properties of a given sample. \citet{Schlaufman:2010fk} became the first to suggest a relation between {spin--orbit} misalignment and a stellar property, but did so without obtaining any observations of the {Rossiter--McLaughlin effect}. The conclusions were reached by noticing that a number of hosts ($> 1.2 M_\odot$) of transiting {hot Jupiter}s have a rotation rate abnormally lower than the general stellar population. This shows that the stellar axis must be preferentially inclined. A similar method is employed by \citet{Mazeh:2015lr} using the amplitude of photometric modulations acquired by {\it Kepler}, to argue that all planet-hosts (to {hot Jupiter}s and other planet types) with $T_{\rm eff}> 6100$K have rotation axes inclined with respect to our line of sight, and therefore, to the orbital plane of their transiting companions.

\runinhead{The alignment of discs.}

Finally, beyond planets, it is of obvious high interest to measure the {spin--orbit angle} between a star and any material surrounding it, like a debris or a protoplanetary disc. Resolving the stellar surface interferometrically, \citet{Le-Bouquin:2009mz} find an alignement between Fomalhaut and its debris ring. 

A collection of debris discs, resolved by {\it Herschel}, show general agreement with discs being coplanar with the inferred inclination for their central star \citep{Watson:2011mz,Greaves:2014vn}.  We will note here the existence of one polar disc surrounding the circumbinary system 99 Herculis \citep{Kennedy:2012zr}, and the presence of a warped disc in  KH 15D \citep{Chiang:2004fk,Winn:2006qy}.  

Misalignments appear more frequent in protoplanetary discs, with large {mutual inclination}s reported between two circumstellar discs surrounding each component of a binary system \citep{Jensen:2014lr}, as well as between circumstellar and circumbinary discs \citep{Brinch:2016lr}. Those however are generally consistent with stellar axes being unrelated to their orbital plane, for separations in excess of 30-40 AU \citep{Hale:1994fk}.

\section{Other uses for the Rossiter-McLaughlin effect}

\runinhead{Stellar binaries.}

The {Rossiter--McLaughlin effect} was devised for binaries prior to its application to planets. However, beyond the 1940s, very few measurements have been obtained. The effect is perceived as a nuisance and generally observers avoid observing close to eclipses.

The fast growth of planetary measurement triggered a regain of interest, with the most striking result presented by \citet{Albrecht:2009fy}, who presented evidence that DI Herculis, a binary system, contains two stars, each with a spin misaligned with their orbital spin.

There are two main projects working along these lines of enquiry. While the BANANA project focuses on near equal mass, high-mass binaries \citep{Albrecht:2011bs}, the EBLM project \citep{Triaud:2013lr,Triaud:2017yu} is concerned with unequal pairs, composed of FGK primaries with secondaries that have masses $< 0.3 M_\odot$. These low mass stars have radii and effective temperatures comparable to many {hot Jupiter}s \citep{Triaud:2014kq}. They mimic a {transit} of a Jupiter-sized planet exactly. As such they provide a natural comparison sample to the {hot Jupiter} population, amongst others for the measure of the {Rossiter--McLaughlin effect} \citep{Triaud:2013lr}.  A similar observation was carried out to verify that Kepler-16B orbits coplanar with Kepler-16A's equator \citep{Winn:2011wd}. This matches the coplanarity of the circumbinary planet Kepler-16AB\,b \citep{Doyle:2011vn}.

\runinhead{Exomoons.}

An {exomoon} has yet to be discovered (see chapter by Heller, R. ``Detecting and Characterizing Exomoons and Exorings''). However, the ubiquity of moons in the Solar system indicates that this is probably only a matter of time. \citet{Zhuang:2012zl} provide a first analysis on what type of a {Rossiter--McLaughlin effect} should be expected while a moon transits a star, following its companion planet. They show that multiple events can yield the orbital inclination of the moon to the planet's orbital plane. 

Another method is likely possible, following the recent successes at resolving planets spectroscopically either from their reflected or their emission spectrum. \citet{Snellen:2014kx} not only resolve $\beta$\,Pictoris\,b, but also measure it $v \sin i_{\rm p}$. For comparable rotation rates to a star, an {exomoon} the size of the Earth would cast a {Rossiter--McLaughlin effect} with an amplitude similar to a {hot Jupiter} transiting a Sun-like star. Other alternative methods such as {Doppler tomography} are also available and are probably preferable.

\runinhead{Planetary spin.}

While the planetary orbital spin may be aligned with the stellar rotation spin, this may not necessarily be the case for the planetary spin (e.g. the Earth, whose spin is inclined by $\sim 23.4^\circ$ compared to its orbital plane, giving rise to seasonal variations). At occultation, we can reconstruct the reflected and emitted spectrum of a planet. As the planet disappears behind the star, parts of its surface are scanned, and the {spin--orbit angle} of the planet to its orbit can in principle be measured although none have been reported so far \citep{Nikolov:2015zl}.

\runinhead{Atmospheric investigations.}

Opacity sources within a planetary atmosphere (atoms, molecules, particles) lead to a variation in apparent size of the planet, which is observed as a variation of the {transit} depth, as a function of wavelength \citep{Seager:2010kx}. Following equation~\ref{eq:firstorder} but replacing $D$ by $\Delta D$, a change in {transit} depth, we can see that those would lead to a variation of the amplitude of the {Rossiter--McLaughlin effect}. This was first proposed, and applied by \citet{Snellen:2004qy} who confirmed an enhanced {Rossiter--McLaughlin effect} on the Sodium D lines for HD\,209458\,b. This has also been observed in the case of WASP-17\,b \citep{Wood:2011lr}, and recently \citet{Di-Gloria:2015qy} reported a marginal detection of the scattering slope of HD\,189733\,b. For these atmospheric observation, the {Rossiter--McLaughlin effect} has to be taken into account to obtain proper results \citep{Louden:2015fk,Brogi:2016xy,Wyttenbach:2017xy}.

An important aspect is that $v \sin i_\star$ acts as an amplifier of atmospheric features. For instance, would TRAPPIST-1\,b \citep{Gillon:2016gh,Gillon:2017yq} have a hydrogen-rich atmospheres \citep[which it does not;][]{de-Wit:2016fk}, spectral features would be expected to cause Rossiter-McLaughlin amplitude variations of several metres per second thanks to the high spin rate of its host \citep{Cloutier:2016fj}.

\runinhead{Differential rotation.}

Like a photometric {transit}, the {Rossiter--McLaughlin effect} can be used to study the star itself. At a near polar orbit, an exoplanet will scan different stellar latitudes. Latitudinal {differential rotation} will affect the shape of the {Rossiter--McLaughlin effect}. Although this effect was deemed too weak \citep[e.g.][]{Triaud:2009qy} to be detected, the development of the {\it reloaded} method allowed a detection on HD\,189733 \citep{Cegla:2016qy}. This opens a new chapter of investigations.

\runinhead{Transit identification.}

Identifying that a planet is transiting is not always straightforward. The {transit} can last a time comparable to a night, or the star can be so bright it has no reference star within the field of view of a ground-based telescope. In those instances the {Rossiter--McLaughlin effect} can help make a detection. Such an observation was carried out on HD\,80606\,b (although it was accompanied by a photometric timeseries \citep{Moutou:2009qv} \citep[read also Chapt. by F. Bouchy ÒHD189733b: The transiting hot Jupiter that revealed a hazy and cloudy atmosphereÓ, ][]{Bouchy:2005lr}. This also provided some of the first hints of significant {spin--orbit} misalignment, which was later confirmed in \citet{Hebrard:2010fu}. Another similar attempt was produced in the case of HD\,156846\,b, a 360d, eccentric planet. The HARPS measurements show no evidence of a {transit}, but these data were only presented in a thesis \citep{Triaud:2011qy}.

Some planets will have a {Rossiter--McLaughlin effect} with an amplitude higher than its Doppler reflex motion. A careful analysis of residuals during {transit} can reveal that the planet is transiting straight away. This might become relevant for Doppler surveys of rapidly rotating late M-dwarfs \citep{Cloutier:2016fj}.

\runinhead{The Rossiter-McLaughlin effect on the Sun.}

During the last {transit} of Venus, \citet{Molaro:2013lr} used HARPS to recover a {Rossiter--McLaughlin effect}, from Sun light reflected off the Moon. These observations were carried out with the intent of replicating conditions close to what happens when collecting photons from a distant star while an exoplanet {transit}s. The effect is tiny with a semi-amplitude of just 1 m s$^{-1}$\citep{Molaro:2013lr}. They demonstrate that in principle, signals as weak as this can be properly recovered, with currently available technology.

The {Rossiter--McLaughlin effect} has also been captured during lunar eclipse, using HARPS \citep{Yan:2015xy}, and using a solar telescope and a Fourier transform spectrograph \citep{Reiners:2016lr}. The effect is stunning with a whooping 1.4~km~s$^{-1}$ total amplitude. These type of observations can be used as a benchmark to investigate multiple stellar effects, notably convective blue-shift and the modelling of centre-to-{limb darkening}.\\

\section{Conclusions }

In the past ten years, we have discovered that the measure of the {spin--orbit} angle $\lambda$ does not provide the clean diagnostics for planetary migration that we had been hoping for. Instead Nature has shown many surprises. These make the interpretation of the {spin--orbit} distribution challenging, but also, so very interesting. The measure of the {spin--orbit angle}, like for many other observable quantities, participates in building a context for each planet.

The future is bright. The first attempts to determine the {spin--orbit} angle of a super-Earth have been made \citep[55 Cnc e;][]{Lopez-Morales:2014qv,Bourrier:2014lr}. They offer different conclusions but those should soon be resolved by up-coming instrument like ESPRESSO, on the VLT (see Gonz‡lez-Hern‡ndez, J., Pepe, F., Molaro, P., Santos, N., ``ESPRESSO on VLT: An Instrument for Exoplanet Research''). High on the agenda is to measure the angle for multiplanetary systems orbiting stars with effective temperatures greater than 6100K, and verify the results provided by \citet{Mazeh:2015lr}. They indicate that super-Earths ought to be found on inclined orbits, supporting the idea that the more massive stars have unstable stellar axes. As TESS \citep[see chapter by see Ricker, G., ``Space Missions for Exoplanet Science: TESS'', and ][]{Sullivan:2015uo} and PLATO \citep[see chapter by Heras, A., Rauer, H., ``Space Missions for Exoplanet Science: PLATO'', and][]{Rauer:2014qq} provide a large sample of super-Earths transiting relatively bright stars, we will have many investigations to carry out.

High-resolution spectrographs are being used more and more regularly for the study of planetary atmospheres, using ground-based facilities. Those observations are obtained at {transit} and contain the {Rossiter--McLaughlin effect}. There are two scientific uses for a same observation. Atmospheric work often required multiple epochs to build up the signal or quantify astrophysical systematics. We will gain a chance to improve the precision we obtain on $\lambda$ and verify that epoch upon epoch provides reproducible results, an important element of the scientific method.

\begin{acknowledgement}
The author would like to acknowledge the many discussions with friends, colleagues, collaborators and competitors that across the years have shaped my thoughts. Notably amongst them are: Simon Albrecht, David Anderson, Adrian Barker, Vincent Bourrier, David Brown, Heather Cegla, Andrew Collier Cameron, Bekki Dawson, Dan Fabrycky, Guillaume H\'ebrard, Dong Lai, Doug Lin, Claire Moutou, Soko Matsumura, Michel Mayor, Norio Narita, Smadar Naoz, Gordon Ogilvie, Cristobal Petrovich, Didier Queloz, Joshua Winn and Yanqin Wu. Finally, the author would like to thank Alexander von Boetticher for proof reading the document, and Roi Alonso for the opportunity to write this chapter.
\end{acknowledgement}

\bibliographystyle{spbasicHBexo}  
\bibliography{Triaud_RM} 

\end{document}